\documentclass[prl,twocolumn,showpacs,preprintnumbers,amsmath,amssymb,superscriptaddress]{revtex4}

\usepackage{bm}
\usepackage{graphicx}
\usepackage{amsbsy}
\usepackage{amsmath}
\usepackage{amsfonts}
\usepackage{amsthm}

\begin{document}

\theoremstyle{plain}
\newtheorem{theorem}{Theorem}
\newtheorem{lemma}[theorem]{Lemma}
\newtheorem{corollary}[theorem]{Corollary}
\newtheorem{conjecture}[theorem]{Conjecture}

\theoremstyle{definition}
\newtheorem{definition}{Definition}

\theoremstyle{remark}
\newtheorem*{remark}{Remark}
\newtheorem{example}{Example}

\title{Deterministic Entanglement of Assistance and Monogamy Constraints}   
\author{Gilad Gour}\email{ggour@math.ucsd.edu}
\affiliation{Department of Mathematics, University of California/San Diego, 
        La Jolla, California 92093-0112}
\author{David A. Meyer}
\affiliation{Department of Mathematics, University of California/San Diego, 
        La Jolla, California 92093-0112}
\author{Barry C. Sanders}
\affiliation{Institute for Quantum Information Science, University of Calgary, Alberta T2N 1N4, Canada}

\date{\today}

\begin{abstract} 
Certain quantum information tasks require entanglement of assistance,
namely a reduction of a tripartite entangled state to a bipartite entangled state via local measurements.
We establish that `concurrence of assistance' (CoA)  identifies capabilities and limitations 
to producing pure bipartite entangled states from pure tripartite entangled states
and prove that CoA is an entanglement monotone for $(2\times2\times n)$-dimensional
pure states. Moreover, if the CoA for the pure tripartite state
is at least as large as the concurrence of the desired
pure bipartite state, then the former may be transformed to the latter
via local operations and classical communication, and we calculate the maximum probability
for this transformation when this condition is not met.
\end{abstract} 


\maketitle

Entanglement is crucial for many quantum information processing tasks. More specifically
bipartite entanglement underpins ubiquitous tasks such as quantum teleportation,
entanglement swapping, and remote state preparation. Three alternatives to producing
bipartite entanglement include processing a product state of two qubits through a 
two-qubit unitary transformation such as a CNOT gate, creating a source of
two-qubit entanglement (such as parametric down conversion for producing
polarization-entangled two-photon states), and the reduction of a multipartite entangled
state to an entangled state over fewer parties (e.g.\ bipartite) via measurements.
Each process is important and the choice of which process should be used depends on
the physics of the implementation.
The latter case, which we call ``assisted entanglement'', quantified by the
entanglement of assistance (EoA)~\cite{DiV98}, is especially important for 
quantum communication, where quantum repeaters are needed to establish bipartite
entanglement over a long length scale~\cite{Bri98}, and  for spin systems that are all coupled via
Ising or similar interactions resulting in a multipartite entangled state~\cite{Bri01}
(which is the resource for one-way quantum computation~\cite{Rau01}).

An important application of assisted entanglement concerns the deterministic creation 
of a bipartite entangled state from a tripartite entangled state, where the tripartite state
consists of three parties: the two qubits to be prepared in a bipartite entangled state and
a party in an $n$-dimensional Hilbert space that corresponds to all other particles in the system.
Here we fully assess deterministic creation of a bipartite entangled state from a 
pure tripartite entangled state, including 
(i)~proving that concurrence of assistance (CoA)~\cite{Lau01} is
 an entanglement monotone, (ii)~presenting
a condition for CoA which, if met, guarantees that deterministic distillation
of a bipartite pure state from a single copy of a tripartite state can be achieved, (iii)~calculating
the maximum probability for obtaining the desired bipartite state 
if this condition is not satisfied, and (iv)~showing that CoA satisfies 
monogamy constraints that are dual to the Coffman-Kundu-Wootters (CKW)
monogamy constraints for concurrence~\cite{Cof00} and recently proven in the general case
by Osborne~\cite{Osb05}. Thus we have provided a strong foundation to analyzing
assisted capabilities and limitations of assisted entanglement, which is an important tool
for creating bipartite entanglement in certain important physical systems.

We now consider a tripartite pure state shared between three parties referred to 
as Alice, Bob, and Sapna: the entanglement supplier, Sapna,
performs a measurement on her share of the
tripartite state, which yields a known bipartite entangled state for Alice+Bob.
Specifically we study a pure  $(2\times 2\times n)$-dimensional tripartite 
entangled state, $|\psi\rangle_\text{ABS}\in\mathcal{H}_2\times\mathcal{H}_2\times\mathcal{H}_n$, 
with $n\geq 2$ the dimension of Sapna's system.
Tracing over Sapna's system yields the bipartite $(2\times 2)$-dimensional mixed state
$\rho_\text{AB}\equiv\text{Tr}_\text{S}\left(|\psi\rangle_\text{ABS}\langle\psi|\right)$
shared by Alice+Bob, and any decomposition
of $\rho_\text{AB}$ can be realized by a generalized measurement performed by Sapna~\cite{Hug93}.

Sapna's aim is to maximize entanglement for Alice+Bob, and the
maximum average entanglement she can create is the EoA, which was
originally defined in terms of entropy
of entanglement (dual to the entanglement of formation)~\cite{DiV98,Coh98}
but extended here to include any bipartite entanglement measure:
the EoA~$E_\text{a}$ for a tripartite pure state $|\psi\rangle_\text{ABS}$ 
is dual to any entanglement measure~$E$ according to
\begin{equation}
	E_\text{a}(|\psi\rangle_\text{ABS})\equiv E_\text{a}(\rho_\text{AB})\equiv 
	\text{max}\sum_k p_k E(|\phi_k\rangle_\text{AB})\;, 
\label{eq:def}
\end{equation}
which is maximized over all possible decompositions of 
$\rho_\text{AB}=\sum_k p_k|\phi_k\rangle_\text{AB}\langle\phi_k|$. 
\begin{remark}
In general, a distribution of states that maximizes Eq.~(\ref{eq:def})
for a given entanglement measure~$E$ will not necessarily be the optimal distribution
for a different measure. Therefore, the choice of measure is important
and depends on the planned quantum information task by Alice and Bob subsequent to
Sapna's assistance.
\end{remark}
For any choice of entanglement monotone~$E$, EoA is bounded, namely
\begin{equation}
E_\text{a}(|\psi\rangle_\text{ABS})\leq \text{min}\{E_\text{A(BS)},E_\text{B(AS)}\}\;,
\label{bound}
\end{equation}
for $E_\text{A(BS)}$ ($E_\text{B(AS)}$) the bipartite entanglement 
shared by Alice with Bob+Sapna (Bob with Alice+Sapna).
Eq.~(\ref{bound}) holds because both
bipartite entanglements~$E_\text{A(BS)}$ and~$E_\text{B(AS)}$ cannot increase 
by any general local operations and classical communications (LOCC) by Alice, Bob, and Sapna. 
This bound is not tight in general,
and tighter bounds have been obtained (e.g.~\cite{Gou04,Gou05}).
Recently, it has been shown~\cite{Smo05} (see also~\cite{Hor05}
for the generalization of this result) that in the asymptotic limit the upper bound in
Eq.~(\ref{bound}) is saturated~\cite{footnote:asymptotic}.

In general the EoA~(\ref{eq:def}) is difficult to calculate, in contrast to the
CoA for the $2\times 2\times n$ pure state $|\psi\rangle_\text{ABS}$~\cite{Lau01}:
\begin{equation}
	C_\text{a}(|\psi\rangle_\text{ABS})=F\left(\rho_\text{AB},\tilde{\rho}_\text{AB}\right),
\label{t}
\end{equation}
with $\tilde{\rho}_\text{AB}$ defined by the ``spin flip'' transformation~\cite{Woo98}
$\tilde{\rho}_\text{AB}=\sigma_y\otimes\sigma_y\rho_\text{AB}^{*}\sigma_y\otimes\sigma_y$
and $F(\rho,\sigma)=\text{Tr}\sqrt{\sqrt{\sigma}\rho\sqrt{\sigma}}$ the fidelity.
Although EoA (including CoA) is not
a bipartite entanglement monotone~\cite{footnote:separable}, 
we consider whether EoA can be an entanglement monotone
for pure tripartite states. In general, it is not known if the EoA~(\ref{eq:def}) provides the maximum average of entanglement under {\em general} three-party LOCC (i.e.\ not only
by Sapna's measurement) and hence may not be an entanglement monotone, but, 
in the following theorem, we prove that CoA is indeed an entanglement monotone.  
\begin{theorem}
	CoA is an entanglement monotone for $2\times 2\times n$ pure states.
\end{theorem}
\begin{proof} 
We need to prove that general LOCC by Alice, Bob and Sapna
cannot increase the CoA for a tripartite $2\times2\times n$ pure state
$|\psi\rangle_\text{ABS}$. Thus, let us consider the following three-way LOCC.
First, Sapna performs a measurement represented by Kraus operators $\hat{M}^{(j)}$ 
and sends result $j$ to Alice+Bob. Based on this result, Alice performs a 
measurement represented by Kraus operators $\hat{A}^{(k)}_j$ and transmits
her result~$k$ to Bob+Sapna. Based on outcomes $j,k$ from Sapna+Alice,
Bob performs a measurement represented by Kraus operator $\hat{B}^{(n)}_{jk}$
and send his result~$n$ to Sapna. Finally Sapna performs
a second measurement with Kraus operator denoted by $\hat{F}_{jkn}^{(i)}$ and sends
the result $i$ to Alice and Bob. 

The final distribution of entangled states shared between Alice and Bob is denoted by 
$\{N_{jkni},|\phi_{jkni}\rangle _{ABS}\}$, where $N_{jkni}$ is the 
probability for outcome $j,k,n,i$. The state $|\phi_{jkni}\rangle _{ABS}$ is given by
\begin{equation}
	|\phi_{jkni}\rangle_\text{ABS}
		=\frac{\hat{A}_{j}^{(k)}\otimes\hat{B}_{jk}^{(n)}
			\otimes\hat{F}_{jkn}^{(i)}\hat{M}^{(j)}}{\sqrt{N_{jkni}}}|\psi\rangle_\text{ABS}\;,
\end{equation}
and the final average of concurrence is given by 
$\langle C_{AB}\rangle\equiv \sum_{jkni}N_{jkni}C(\sigma_\text{AB}^{jkni})$
for
$\sigma _{AB}^{jkni}\equiv \text{Tr} _{S}|\phi_{jkni}\rangle_\text{ABS}\langle\phi_{jkni}|$.
As the concurrence of any bipartite state $|\phi\rangle_\text{AB}$
satisfies 
$C(\hat{A}_{j}^{(k)}\otimes\hat{B}_{jk}^{(n)}|\phi\rangle)=|\text{Det}(\hat{A}_{j}^{(k)})|
|\text{Det}(\hat{B}_{jk}^{(n)})|\;C(|\phi\rangle)$, we obtain
\begin{align}
\langle C_{AB}\rangle & = 
\sum_{jkni}|\text{Det}(\hat{A}_{j}^{(k)})|\;
|\text{Det}(\hat{B}_{jk}^{(n)})|\nonumber\\
& \times C\left(\text{Tr}_{S}
\left[\hat{F}^{(i)}_{jkn}\hat{M}^{(j)}
|\psi\rangle_\text{ABS}\langle\psi|
\hat{M}^{(j)\dag}\hat{F}^{(i)\dag}_{jkn}\right]\right)\nonumber\\
& \leq \sum_{jkn}|\text{Det}(\hat{A}_{j}^{(k)})|\;
|\text{Det}(\hat{B}_{jk}^{(n)})|C_\text{a}\left(
\hat{M}^{(j)}
|\psi\rangle_\text{ABS}\right)\nonumber\\
& \leq \sum_{j} C_\text{a}\left(\hat{M}^{(j)}
|\psi\rangle_\text{ABS}\right)\leq C_\text{a}\left(
|\psi\rangle_\text{ABS}\right)\;.
\end{align}
The first inequality follows from the fact that the second measurement performed by Sapna
(represented by the Kraus operators $\hat{F}^{(i)\dag}_{jkn}$) yields a probability
distribution of states with average concurrence smaller than the concurrence of assistance
of $\hat{M}^{(j)}|\psi\rangle_\text{ABS}$. The second inequality follows from
the geometric-arithmetic inequality: 
$\sum_{n}\left|\text{Det}(\hat{B}_{jk}^{(n)})\right|\leq 
\tfrac{1}{2}\sum _{n}\text{Tr}\hat{B}_{jk}^{(n)\dag}\hat{B}_{jk}^{(n)}=1$ (and similarly
for $\hat{A}_j^{(k)}$).
Evidently all operations that are performed by Alice, Bob and Sapna
cannot yield a probability distribution with average concurrence (between Alice+Bob)
that exceeds the CoA. 
\end{proof}  

CoA is a readily computed measure of entanglement for tripartite systems 
that can serve as a convenient mathematical tool to determine when a transformation
between two tripartite states cannot be realized by LOCC. 
\begin{example}
Consider two tripartite states 
$|\psi\rangle=\sqrt{a}|100\rangle+\sqrt{b}|010\rangle+\sqrt{c}|001\rangle$ and
$|\phi\rangle=\sqrt{a}|100\rangle+\sqrt{c}|010\rangle+\sqrt{b}|001\rangle$ with
$a+b+c=1$ and $b > c$ (both states belong to the W-class~\cite{Dur00}). Let  
$C_\text{a}^k $ ($k=1,2,3$) be the CoA with respect to
Sapna, who holds the $k^\text{th}$ qubit. 
From Eq.~(\ref{t}), we obtain $C_\text{a}^{3}(\psi)\geq C_\text{a}^{3}(\phi)$, but 
$C_\text{a}^1 (\psi)\leq C_\text{a}^1 (\phi)$.
As both $C_\text{a}^1 $ and $C_\text{a}^{3}$ are entanglement monotones, 
deterministic LOCC transformations between $|\psi\rangle$ and $|\phi\rangle$ cannot
be realized. 
\end{example}

In the following theorem we establish a sufficient condition, based on CoA,
which indicates
if a transformation from tripartite state to bipartite state can be realized deterministically 
by LOCC.
   
\begin{theorem}
A deterministic map
$T:\mathcal{H}_2\times\mathcal{H}_2\times\mathcal{H}_n\rightarrow\mathcal{H}_2\times\mathcal{H}_2:|\psi\rangle_\text{ABS}\mapsto |\phi\rangle_\text{AB}$ can be realized by LOCC iff
\begin{equation}
	C_\text{a}(|\psi\rangle_\text{ABS})\geq C(|\phi\rangle_\text{AB}).
\label{eq:tt}
\end{equation}
\label{theorem:deterministicmap}
\end{theorem}

\begin{proof}
Inequality~(\ref{eq:tt}) is a necessary condition because CoA
is an entanglement monotone; thus the onus is now
to prove that inequality~(\ref{eq:tt}) is a sufficient condition.

Beginning with the decomposition $\rho_\text{AB}=\sum_{k=1}^{n}|\phi_k\rangle\langle\phi_k|$
for $n\leq 4$ the rank of $\rho_\text{AB}$ and $|\phi_k\rangle$ 
subnormalised such that 
$\langle\phi_k|\tilde{\phi}_{k'}\rangle =\lambda_k \delta_{kk'}$ (see~\cite{Woo98}),
we observe that the average concurrence of this decomposition is $\sum_{k=1}^n\lambda_k$,
which is optimal; c.f.\  Eq.~(\ref{t}). Any other decomposition of 
$\rho_\text{AB}=\sum_{l=1}^m |\chi_l\rangle\langle\chi_l|$ is given by
$|\chi_l\rangle=\sum_{k=1}^{n}U_{lk}^{*}|\phi_k\rangle$
with $m\geq n$ and $U$ an $m\times m$ unitary matrix.
Thus, the average concurrence of the decomposition, $\chi$, is given by
	$\langle C\rangle=\sum_{l=1}^m \left|\sum_{k=1}^{n}\left(U_{lk}\right)^2 \lambda _k\right|$.
Thus, any other decomposition, $\chi$, has the same average concurrence 
as $\phi$ as long as the matrix elements $U_{lk}$ are all real
(i.e.\ $U$ is an $m\times m$ orthogonal matrix).
Hence, as discussed in~\cite{Woo98}, one can always find an optimal decomposition such that
all states in the decomposition have the same concurrence. According
to Eq.~(\ref{eq:tt}) 
this concurrence exceeds $C(|\phi\rangle_\text{AB})$ so, since all bipartite states
in this decomposition
have dimension $2\times 2$, they are all majorized by $|\phi\rangle_\text{AB}$, and it follows 
from Nielsen's theorem~\cite{Nie99} that Alice, Bob and Sapna can 
transform $|\psi\rangle_\text{ABS}$ to $|\phi\rangle_\text{AB}$ by LOCC.
\end{proof}

In addition to CoA being valuable for testing whether deterministic LOCC transformations
map from a pure tripartite state to tripartite or bipartite states, the CoA 
for the $2\times2\times2$ pure state $|\psi\rangle_\text{ABS}$ also exhibits
monogamy constraints~\cite{Osb05}
(entanglement tradeoffs) analogous to those for the usual concurrence.
Here we derive another monogamy constraint which is in some sense the {\it dual}
to the CKW constraint~\cite{Cof00,Osb05}.

\begin{theorem}
For a pure tripartite state in $\mathcal{H}_2\times\mathcal{H}_2\times\mathcal{H}_2$
and $\tau_{ABS}\equiv C^2_\text{A(BS)}-C^2 (\rho_\text{AB})-C^2 (\rho_\text{AS})$ 
the 3-tangle~\cite{Cof00} with
$\rho_\text{AB}$ and $\rho_\text{AS}$ the reduced density matrices
after tracing over Sapna's and Bob's systems, respectively, 
\begin{equation}
	\tau_{ABS}=C^2_\text{a}(\rho_\text{AB})+C^2_\text{a}(\rho_\text{AS})-C^2_\text{A(BS)}\geq 0.
\label{eq:tau}
\end{equation}
\end{theorem} 
\begin{proof} 
We employ CKW's notation, wherein $\lambda_1^\text{AB}$, $\lambda_2 ^\text{AB}$ 
and $\lambda_1^\text{AS}$, $\lambda_2 ^\text{AS}$ denote the eigenvalues of
$R_{AB}\equiv\sqrt{\sqrt{\rho_\text{AB}}\tilde{\rho}_\text{AB}\sqrt{\rho_\text{AB}}}$ and 
$R_\text{AS}\equiv\sqrt{\sqrt{\rho_\text{AS}}\tilde{\rho}_\text{AS}\sqrt{\rho_\text{AS}}}$, respectively.
CKW have shown that $\text{Tr}(\rho_\text{AB}\tilde{\rho}_\text{AB})
+\text{Tr}(\rho_\text{AS}\tilde{\rho}_\text{AS})=C^2_\text{A(BS)}$; therefore,
\begin{align}
&C^2_\text{a}(\rho_\text{AB})+C^2_\text{a}(\rho_\text{AS})-C^2_\text{A(BS)}\nonumber\\
&=(\text{Tr}R_{AB})^2 +(\text{Tr}R_\text{AS})^2-\text{Tr}R_{AB}^2 
-\text{Tr}R_\text{AS}^2  \nonumber\\
& = 2( \lambda_1^\text{AB}\lambda_2 ^\text{AB}+\lambda_1^\text{AS}\lambda_2 ^\text{AS}).
\end{align}
CKW have shown that  
$\tau_{ABS}=2( \lambda_1^\text{AB}\lambda_2 ^\text{AB}+\lambda_1^\text{AS}\lambda_2 ^\text{AS})$,
which proves the theorem.
\end{proof}

CKW conjectured~\cite{Cof00} (recently proven by Osborne~\cite{Osb05})
that, for $n$ qubits (labeled by $1,2,...,n$),
\begin{equation}
	C^2(\rho_{12})+C^2(\rho_{13})+\cdots +C^2(\rho_{1n})\leq C_{1(23...n)}^2 \;.
\label{00}
\end{equation}
Similary, we are willing to conjecture that the dual to this conjecture also holds.
\begin{conjecture}
\begin{equation}
	C_\text{a}^2(\rho_{12})+C_\text{a}^2(\rho_{13})+\cdots +C_\text{a}^2
	(\rho_{1n})\geq C_{1(23...n)}^2 .
\label{11}
\end{equation}
\end{conjecture}
It is interesting to note that for states of the form
\begin{equation}
|\phi\rangle=\alpha_1|10...0\rangle+\alpha_2|010...0\rangle+\cdots\alpha_n|0...01\rangle\;, 
\end{equation}
both inequalities~(\ref{00},\ref{11}) become equalities.

In Theorem~\ref{theorem:deterministicmap} we established a necessary and sufficient
condition for the existence of a deterministic transformation from a pure 
$2\times2\times n$ tripartite state to a bipartite state. When this condition is violated,
such a transformation may be possible but only probabilistically. 
Here we consider a $d\times d\times n$ pure tripartite state 
$|\psi\rangle_\text{ABS}$ shared by Alice, Bob and Sapna and investigate the maximum probability,
$\mathcal{P}_m$, to `distill' a bipartite state $|\phi\rangle_\text{AB}$ according to the
protocol that Sapna first performs a generalized measurement, whose outcome is communicated
to Alice and Bob, and Alice and Bob subsequently perform pairwise LOCC.

The maximum probability, $P_m$, to transform locally one $d\times d$ bipartite state to another
is given by~\cite{Vid98}:
\begin{equation}
P_m(|\psi\rangle\rightarrow|\phi\rangle)
	=\text{min}\left\{\frac{E_1(|\psi\rangle)}{E_1(|\phi\rangle)},\ldots,
		\frac{E_d(|\psi\rangle)}{E_d(|\phi\rangle)}\right\}
\end{equation} 
for $E_l(|\psi\rangle)=\sum_{k=l}\lambda_k$, where $\lambda_k$ are the Schmidt numbers of 
$|\psi\rangle$ in a decreasing order. NB: for a given fixed state $\phi$, 
$E_{\phi}(|\psi\rangle)\equiv P_m(|\psi\rangle\rightarrow|\phi\rangle)$ is an entanglement 
monotone.
The probability to `distill' the bipartite state $|\phi\rangle_\text{AB}$
from $|\psi\rangle_\text{ABS}$ is, therefore, given by
\begin{equation}
\mathcal{P}_m(|\psi\rangle_\text{ABS}\rightarrow|\phi\rangle_\text{AB})
=\text{max}\sum_{i}p_iE_{\phi}(|\psi _i\rangle_{AB})\;,
\end{equation} 
where the maximum is take over all the decompositions of 
$\rho_\text{AB}=\sum_{i}p_i|\psi_i\rangle_\text{AB}\langle\psi _i|$.
That is, the EoA when measured in terms of the monotone, $E_{\phi}$, can be
interpreted as the maximum probability to distill the bipartite state 
$|\phi\rangle$.
In particular, the maximum probability to distill a $d \times d$ maximally 
entangled state is
\begin{equation}
\mathcal{P}_m=\text{max}\sum_{i}p_i
E_d(|\psi _i\rangle_\text{AB})\;,
\label{mp}
\end{equation}
where the normalized entanglement monotone, $E_d(\psi)\equiv d\lambda_\text{min}(\psi)$,
and $\lambda_\text{min}(\psi)$ is the minimum Schmidt number (including zero) of $\psi$. 
In general, it is quite difficult to calculate $\mathcal{P}_m$, however, in the following 
we calculate it for a large class of $(2\times2\times n)$-dimensional pure states.

\begin{lemma}
For $a_{ij}^l \equiv {}_{A}\langle i|_{B}\langle j|_{S}\langle l|\psi\rangle_\text{ABS}$ the
complex components of 
$|\psi\rangle_\text{ABS}\in\mathcal{H}_2\times\mathcal{H}_2\times\mathcal{H}_n$,
there exists orthonormal bases $|i\rangle_\text{A}$, $|j\rangle_\text{B}$, and 
$|l\rangle_\text{S}$ for Alice, Bob and Sapna, respectively, such that the matrices 
$A^{l\dag}A^l $ are diagonal for {\em all} $l$ with $A^l\equiv(a_{ij})^l$ a $2\times2$
matrix of components;
similarly, there exists orthonormal bases for which $A^l A^{l\dag}$ is diagonal for all $l$. 
\label{lemma:aijl}
\end{lemma}

\begin{proof}
According to the Schmidt decomposition, there are two orthonormal bases $\{|i\rangle_\text{A}\}$ and 
$\{|j\rangle_\text{B}\}$ and two sets of orthonormal states $\{|\chi_{i}\rangle_\text{BS}\}$ and 
$\{|\eta_{j}\rangle_\text{AS}\}$ such that 
\begin{equation}
|\psi\rangle_\text{ABS}=\sum_{i=0}^1 \sqrt{p_i}|i\rangle_\text{A}|\chi_{i}\rangle_\text{BS}
=\sum_{j=0}^1 \sqrt{q_j}|j\rangle_\text{B}|\eta_{j}\rangle_\text{AS}\;,
\label{cen}
\end{equation}
with $p_i$ and $q_j$ the corresponding Schmidt numbers.
With respect to these bases $\{|i\rangle_\text{A}\}$ and $\{|j\rangle_\text{B}\}$
we have $a_{ij}^l =\sqrt{q_j}{}_{A}\langle i|_{S} \langle l|\eta_{j}\rangle_\text{AS}$, where
the orthonormal basis $|l\rangle _{S}$ will be determined later. We now define four
$n\times n$ matrices $\tau^{(j,j')}$ ($j,j'=0,1$) whose components are:
\begin{equation}
	\left(\tau^{(j,j')}\right)_{ll'}
	\equiv\sum_{i=0}^1 a_{ij}^{l*}a_{ij'}^{l'}\;.
\label{eq:taudef}
\end{equation}
Since the states $\{|\eta_{j}\rangle_\text{AS}\}$ are orthonormal, it follows that
\begin{equation}
\text{Tr}\,\tau^{(j,j')}=q_j\delta_{j,j'}\;.
\label{tric}
\end{equation}
Note that the $2\times 2$ matrix $A^{l\dag}A^l $ can be written as
\begin{equation}
A^{l\dag}A^l =
\begin{pmatrix}
\left(\tau^{(0,0)}\right)_{ll} &  \left(\tau^{(0,1)}\right)_{ll}\cr
\left(\tau^{(1,0)}\right)_{ll} & \left(\tau^{(1,1)}\right)_{ll} \cr
\end{pmatrix}\;.
\label{aa}
\end{equation}
Consider a basis change for Sapna's system:
$|l\rangle_\text{S}\rightarrow U|l\rangle_\text{S}$ with~$U$ an $n \times n$ 
unitary matrix. From definitions~(\ref{eq:taudef}), the new $\tau$ matrices are
$U\tau^{(j,j')}U^{\dag}$. As the matrix $\tau^{(0,1)}=\tau^{(1,0)\dag}$ 
has zero trace (c.f.\ Eq.~(\ref{tric})), the theorem by Walgate et al~\cite{Wal00} implies there is a unitary 
matrix such $U\tau^{(0,1)}U^{\dag}$ has zero diagonal. From Eq.~(\ref{aa})
it follows that, for this $U$, all the matrices $A^{l\dag}A^l $ are diagonal,
and similar reasoning for a different basis applies to $A^lA^{l\dag}$.
\end{proof}

We now define a class $\mathcal{A}$ of $2\times 2\times n$ pure states
for which we are able to calculate explicitly the maximum probability to distill 
a Bell state (i.e. Eq.~(\ref{mp}) for $d=2$). We first denote by $E_\text{A(BS)}$ ($E_\text{B(AS)}$)
the {\em bipartite} entanglement (as defined below Eq.~(\ref{mp}) for $d=2$)
between the system of Alice and the joint system Bob+Sapna (Bob and Alice+Sapna).
From Eq.~(\ref{cen}), $E_\text{A(BS)}=2p_1$ and 
$E_\text{B(AS)}=2q_1$ (assuming $p_0\geq p_1$ and $q_0\geq q_1$). 
According to Lemma~\ref{lemma:aijl}, 
there are bases for which the $2\times 2$ matrices 
$A^{l\dag}A^l =\text{diag}\{q_0 ^l ,q_1^l \}$ ($q_0 ^l $, $q_1^l $ denote
eigenvalues), and there are \emph{other} bases for which 
$A'{}^l A'{}^{l\dag}=\text{diag}\{p_0 ^l ,p_1^l \}$ ($A'$ indicates a
different basis: $A^l \neq A'{}^l $). NB:
$q_0=\sum_{l} q_0 ^l $, $q_1=\sum_{l} q_1^l $ and $p_0=\sum_{l} p_0 ^l $, 
$p_1=\sum_{l} p_1^l $.
Define class $\mathcal{A}$:  
$|\psi\rangle _{ABS}\in \mathcal{A}$ if (i)~$E_\text{A(BS)}\geq E_\text{B(AS)}$ and
(ii)~$q_0 ^l \geq q_1^l $
$\forall l$ \emph{or} (i') $E_\text{A(BS)}\leq E_\text{B(AS)}$ and (ii') $p_0 ^l \geq p_1^l $ 
$\forall l$.

The bound in Eq.~(\ref{bound}) saturates for class~$\mathcal{A}$:
\begin{theorem}
If $|\psi\rangle_\text{ABS}\in \mathcal{A}$ then
the maximum probability $\mathcal{P}_m$ is given by
\begin{equation}
\mathcal{P}_m=\text{min}\{E_\text{A(BS)},\;E_\text{B(AS)}\}\;.
\label{t3}
\end{equation}
\end{theorem}
\begin{proof}
The average of $E_2$ (see Eq.~(\ref{mp}) with $d=2$) between Alice and Bob
after Sapna performs the projective basis measurement in the basis $|l\rangle_S$ 
(as defined in Lemma~\ref{lemma:aijl}) gives the 
desired result of Eq.~(\ref{t3}). As $\mathcal{P}_m$ cannot be greater than $\text{min}\{E_\text{A(BS)},\;E_\text{B(AS)}\}$
this is also the optimal result.
\end{proof}
As $E_2\leq E$ for all entanglement measures~$E$ (normalized such that $E=1$ for a Bell state), 
the EoA in terms of $E_2$
provides a lower bound for the EoA when measured with respect to \emph{any} measure $E$.

In the following example the states belong to the class $\mathcal{A}$ 
so Eq.~(\ref{t3}) is correct for these cases.
\begin{example}
\label{ex:W}
The state (a)~$\sqrt{a}|100\rangle+\sqrt{b}|010\rangle+\sqrt{c}|001\rangle$
from the W~class, with $a+b+c=1$, the state (b)~$\sqrt{a}|000\rangle+\sqrt{b}|111\rangle$
from the GHZ~class, with $a+b=1$, and the state
(c)~$|\phi_1\rangle_\text{AS$_1$}|\phi_2\rangle_\text{AS$_2$}
	\in\mathcal{H}_2\times\mathcal{H}_2\times\mathcal{H}_2\times\mathcal{H}_2$
(with Sapna holding the two qubits S$_1$ and S$_2$, which is the case
for entanglement swapping) are in~$\mathcal{A}$.
\end{example}

In conclusion we have proved that CoA is an entanglement monotone that provides a 
condition to assess whether a deterministic transformation exists for creating a bipartite
entangled state from a $2\times2\times n$ pure tripartite state, and we have calculated
the maximum probability for obtaining such states when this condition is not satisfied.
The CoA provides an elegant approach to studying assisted entanglement and obeys
monogamy constraints. Our analysis provides a foundation for studying the capabilities
and limitations of assisted entanglement for producing bipartite entangled states from
multipartite entangled states.

\emph{Acknowledgments:---}
We appreciate valuable discussions with S.~Bandyopadhyay, J.~Oppenheim, A.~Scott,
and J.~Walgate.
G.G.\ and D.A.M.\ acknowledge support by the DARPA QuIST program under
Contract No.\ F49620-02-C-0010 and the NSF under Grant No.\ ECS-0202087,
and B.C.S.\ acknowledges financial support from an Alberta iCORE grant.


\begin{references}
\bibitem {footnote:asymptotic} In the asymptotic limit the EoA is measured 
	in terms of the entropy of entanglement.
\bibitem {footnote:separable} For example the separable state 
	$\openone=\tfrac{1}{4}\sum_{\epsilon,\epsilon'=0}^1
	|\epsilon\epsilon'\rangle\langle\epsilon\epsilon'|$ has one ebit of EoA.
\end{references}
\end{document}